\documentclass{icrc29}
\usepackage{graphicx,amssymb,amsmath,times}
\begin{document}
\title[Impact of a new Cherenkov light parameterisation on the reconstruction of shower profiles from Auger hybrid data]
{Impact of a new Cherenkov light parameterisation on the reconstruction of shower profiles from Auger hybrid data}
\author[F. Nerling et al.] 
{F.~Nerling$^{a}$, J.~Bl\"umer$^{a,b}$, R.~Engel$^a$, M.~Unger$^a$ for the Pierre Auger Collaboration$^c$ 
        \newauthor
\\
        (a) Forschungszentrum Karlsruhe,  Institut f\"ur Kernphysik,
	    76021 Karls\-ruhe, Germany \\ 
        (b) Universit\"at Karlsruhe,  Institut f\"ur Experimentelle Kernphysik,
	    76021 Karls\-ruhe, Germany \\
        (c) for the Auger authors list see these proceedings
        }
\presenter{Presenter: F. Nerling (frank.nerling@ik.fzk.de), \  
ger-nerling-F-abs2-he14-poster}

\maketitle

\begin{abstract}

The light signal measured by fluorescence telescopes receives - strongly depending on 
the shower geometry with respect to the detector - a non-negligible contribution from 
additionally produced Cherenkov light. This Cherenkov contribution has to be accounted for 
to determine primary parameters properly. In comparison to the previous ansatz 
used by other experiments, the impact of a new analytical description of Cherenkov light 
production in EAS on the Auger event reconstruction is investigated. 

\end{abstract}

\section{Introduction}
\begin{figure}[b]
\begin{minipage}[c]{.49\linewidth}
\begin{center}
\includegraphics[clip,bb= 3 20 525 369,width=1.\linewidth]{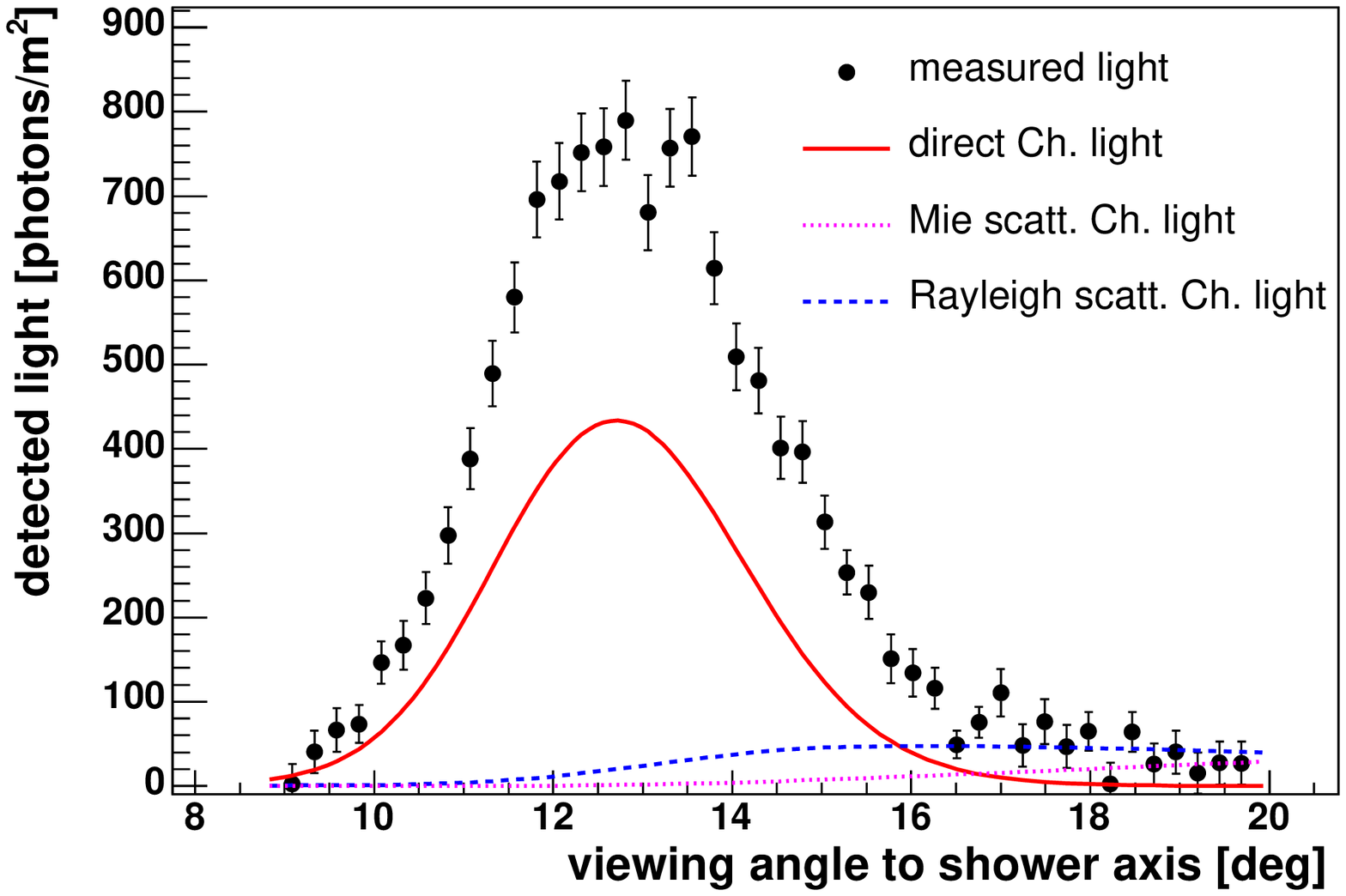}
\end{center}
\vspace{-0.5pc}
\end{minipage}\hfill
\begin{minipage}[c]{.49\linewidth}
\begin{center}
\includegraphics[clip,bb= 3 20 525 369,width=1.\linewidth]{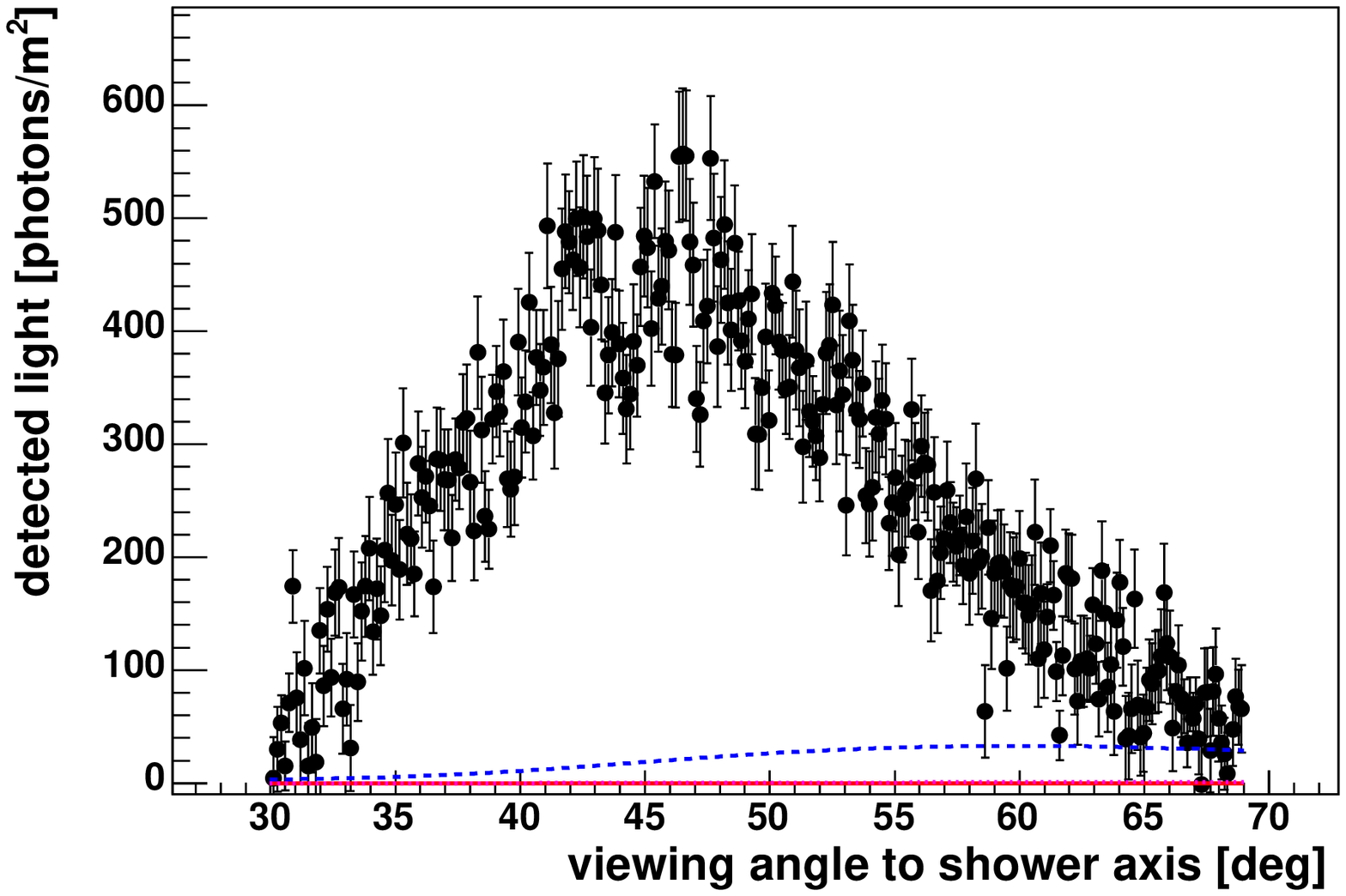}
\end{center}
\vspace{-0.5pc}
\end{minipage}
\caption{Measured light profiles and reconstructed Cherenkov light of an Auger event observed by two
telescopes under different viewing angles (and from different distances).}
\label{fig.recProfile}
\end{figure}
The Pierre Auger Observatory \cite{auger} applies the fluorescence technique for calorimetric measurement 
of longitudinal shower profiles of high-energy EAS.
For the determination of primary parameters based on fluorescence observations, a knowledge 
of the Cherenkov light contribution to the measured light signal is mandatory. 
The amount of Cherenkov light in the fluorescence detector signal depends on the viewing angle with
respect to the shower axis because the Cherenkov photons are emitted mainly in the forward
direction. Due to the steep angular distribution of charged particles in a shower only
at small viewing angles a significant amount of so-called {\it direct Cherenkov light} is
detected. However, direct Cherenkov light can outnumber the fluorescence light by far. 
At larger viewing angles, the Cherenkov light contribution is dominated by photons emitted along 
the shower axis that are scattered into the field of view of the detector. 
This is illustrated in Fig.\,\ref{fig.recProfile}, where a shower simultaneously detected by two 
Auger fluorescence telescopes under different viewing angles is shown. As can be seen, one detector receives a large
amount of Cherenkov light. Clearly, a precise model of Cherenkov light production is needed to infer 
the energy $E$ and position of shower maximum $X_{\rm max}$ of such a shower.
Parameterisations going back to Hillas \cite{hillas 1982} are typically used for calculating analytically the 
Cherenkov light contribution to light signals measured in fluorescence observations, see e.g.
\cite{baltrusaitis 1985,hires}. 
Based on CORSIKA \cite{heck 1998}, QGSJET01 \cite{kalmykov 1997} simulations, a new parameterisation of Cherenkov light production 
providing both the direct and scattered Cherenkov light contributions, has been introduced 
recently \cite{nerling 2005}. 

\section{Reconstructed fraction of Cherenkov light}
Hybrid fluorescence data from 01/04 to 04/05 have been analysed, using the Auger Offline 
\cite{nellen 2005} reconstruction framework, for studying the differences in 
reconstructed event properties due to the new Cherenkov calculation instead of the one 
described in \cite{baltrusaitis 1985}. 
To ensure a reasonable reconstruction quality, the following selection criteria are
applied to the data. 
The $\chi^2/{\rm ndof}$ of the the shower profile fit with a Gaisser-Hillas function is demanded to 
be less than 3 and the estimated statistical relative uncertainties of reconstructed $E$ and 
$X_{\rm max}$ are required to be smaller than 30\,\%. 
The differences in reconstructed event properties $\Delta P_{\rm rec}$ as shown in the following are
always calculated on an event-by-event basis as 
$\Delta P_{\rm rec}=P_{\rm rec}({\rm new}) - P_{\rm rec}({\rm old})$, 
and relative differences are always given relative to the results of the new Cherenkov model. 
The viewing angle under which a single event is seen by the detector changes with shower development.
Therefore, we define an effective viewing angle as the angle between the normal vector of that 
triggered pixel having the mean trigger time with respect to the total shower observation time and 
the shower axis. This mean viewing angle is chosen for the studies presented.
\begin{figure}[t]
\begin{minipage}[c]{.49\linewidth}
\begin{center}
\includegraphics[clip, width=1.\linewidth]{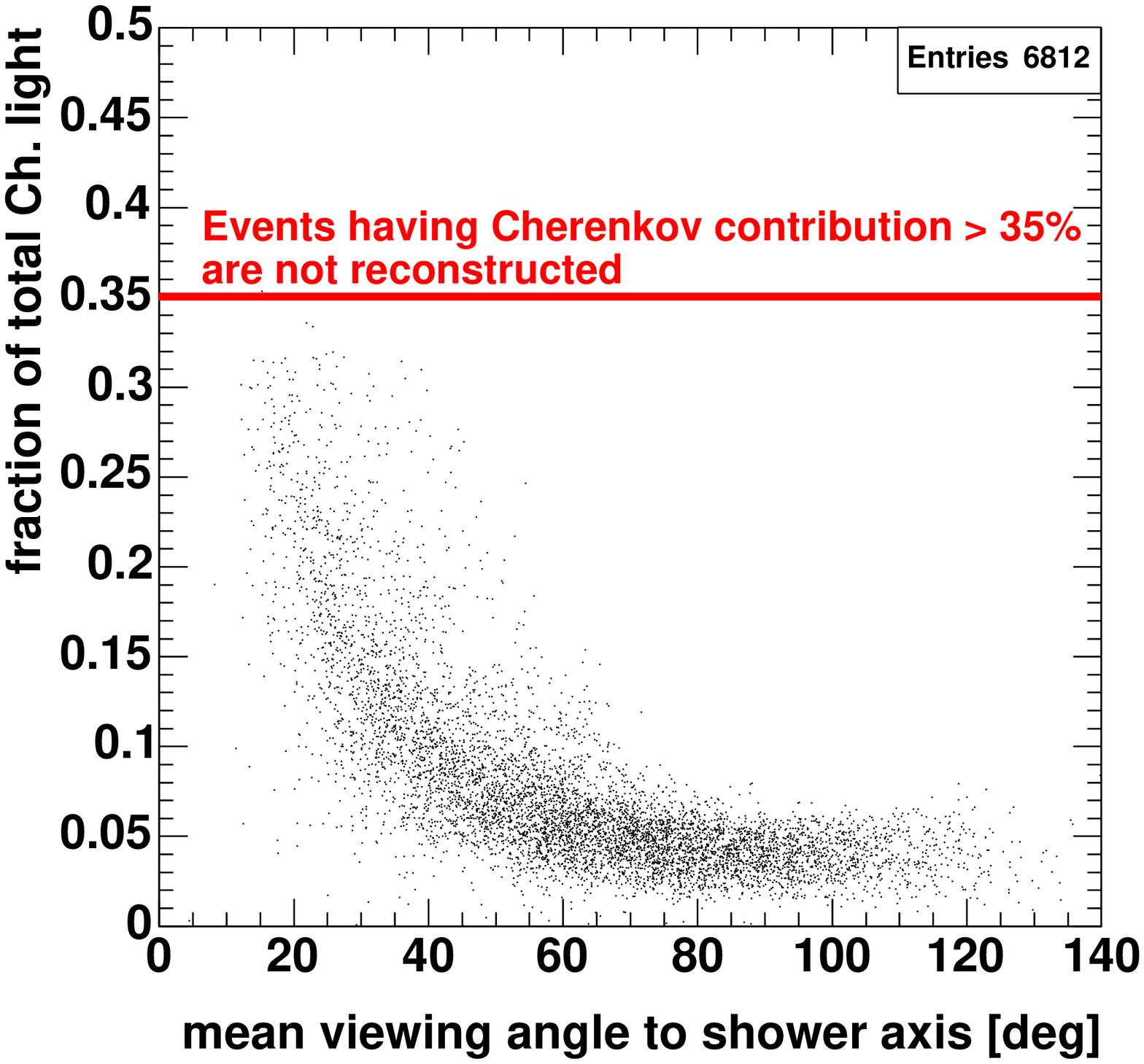}
\caption{Distribution of reconstructed fraction of Cherenkov light contribution to the total shower
signal. \newline ~ \newline ~}
\label{fig.CHfrac_vs_theta}
\end{center}
\vspace{-0.5pc}
\end{minipage}\hfill
\begin{minipage}[c]{.49\linewidth}
\begin{center}
\includegraphics[clip,bb= 27 27 546 506,width=1.\linewidth]{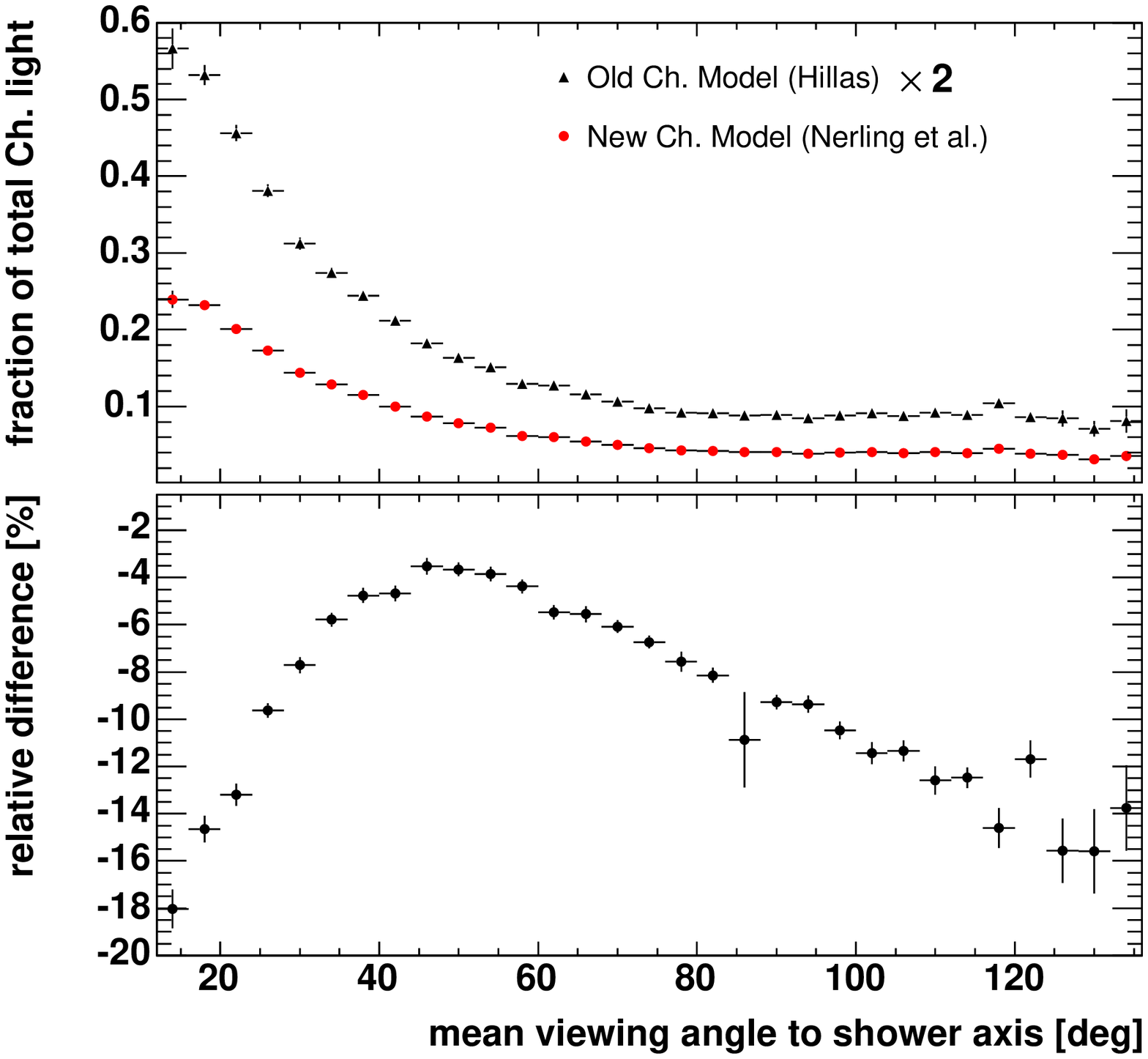}
\caption{Mean fraction of the reconstructed Cherenkov light contribution. 
Data points based on the previous Cherenkov model are scaled by a factor of two.}
\label{fig.CHfrac}
\end{center}
\vspace{-0.5pc}
\end{minipage}
\end{figure}
From Fig.\,\ref{fig.CHfrac_vs_theta} it can be seen that currently only light profiles with less 
than about 35\,\% Cherenkov light are passing the full reconstruction with given quality cuts.
In Fig.\,\ref{fig.CHfrac} the reconstructed fraction of total (superposition of direct and 
scattered) Cherenkov light, defined as $f_{Ch}= N_{\gamma}^{Ch}/(N_{\gamma}^{Ch}+N_{\gamma}^{Fl})$, 
is shown as a function of the mean viewing angle. Here $N_{\gamma}^{Ch}$ is the number
of reconstructed Cherenkov photons and $N_{\gamma}^{Ch}+N_{\gamma}^{Fl}$ the measured light signal
(Cherenkov and fluorescence photons). 
The application of the new model results in smaller fractions of the total Cherenkov light over the whole 
angular range. 
The differences range from about -3\,\% up to -15\,\% at large viewing angles.
For viewing angles smaller than about $40^{\circ}$, the differences increase strongly,
approaching -20\,\%. They are expected to be even larger for viewing angles smaller than
$10^{\circ}$. However, the comparison is hampered in this phase space by the limited statistics. 
It can be concluded that the new model predicts significantly less Cherenkov light depending
systematically on viewing angle. 

\section{Impact on reconstructed primary parameters}
\begin{figure}[t]
\begin{center}
\includegraphics[clip,bb= 21 10 543 524,width=0.65\linewidth]{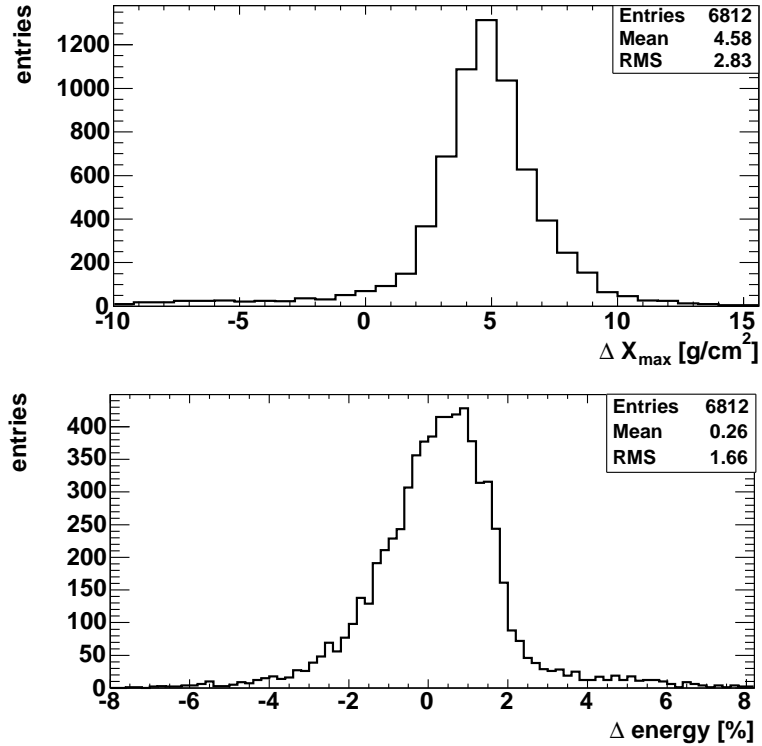}
\end{center}
\caption{Impact of the new Cherenkov calculation on 
reconstructed primary energy and position of shower maximum.}
\label{fig.dEdXmax}
\end{figure}
\begin{figure}[t]

\begin{center}
\includegraphics[clip,bb= 1 3 553 430,width=0.65\linewidth]{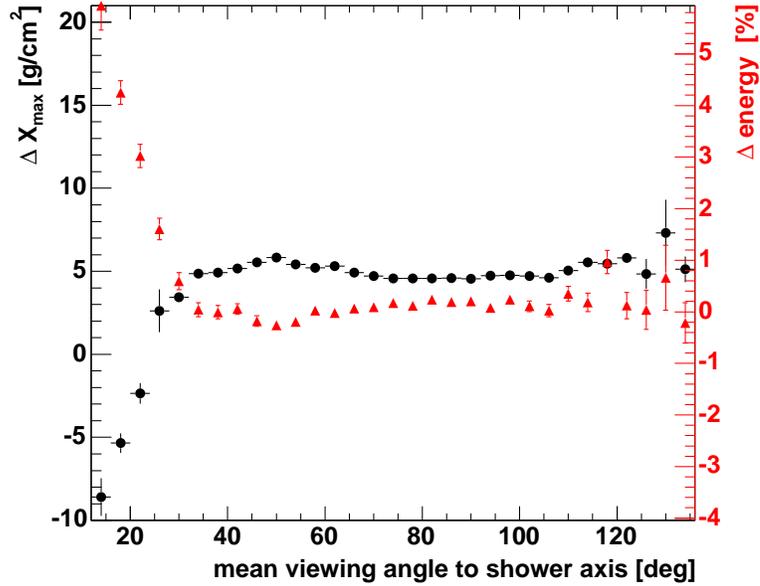}
\end{center}
\caption{Differences of reconstructed energy and position of shower maximum due to the new Cherenkov
calculation as a function of mean viewing angle.}
\label{fig.dEdXmax_2SA}
\end{figure}
The impact of the new Cherenkov calculation on the reconstruction of $E$ and $X_{\rm max}$ 
is shown in Fig.\,\ref{fig.dEdXmax} and Fig.\,\ref{fig.dEdXmax_2SA}. 
The differences in depth of maximum, $\Delta X_{\rm max}$, are given in g/cm$^2$ and energy, 
$\Delta E$, in percent. The mean difference averaged over the complete data set amounts to about 
+5\,g/cm$^{2}$ and +0.3\,\%, respectively. 
As the differences of the reconstructed fraction of Cherenkov light depend on viewing 
angle, a similar dependence is observed for the primary parameters. 
This study is shown in Fig.\,\ref{fig.dEdXmax_2SA}, where $\Delta E$ and  
$\Delta X_{\rm max}$ are given versus the mean viewing angle. 
The difference in reconstructed energy $E$ can
be as large as +6\,\% for smaller angles. The reconstructed depth of maximum increases by about 
+5\,g/cm$^{2}$ for viewing angles larger than about $30^{\circ}$. 
For viewing angles smaller than about $30^{\circ}$ the application of the new Cherenkov calculation 
results in smaller $X_{\rm max}$ values, with the difference increasing up to about -10\,g/cm$^2$ 
at small viewing angles. 
For angles smaller than about $10^{\circ}$, the impact of the new model on inferred primary
parameters is expected to be even larger. However, currently shower profiles comprising Cherenkov 
fractions larger than 35\,\% are rejected in the standard Auger reconstruction.   
\section{Conclusions}
A new Cherenkov light parameterisation \cite{nerling 2005} has been applied for shower profile 
reconstruction of the Auger hybrid data. The reconstructed energy and depth of shower maximum have
been compared to the traditional treatment \cite{baltrusaitis 1985,hires} on an event-by-event basis. 
The resulting differences are significant and depend systematically on the viewing
angle. Events detected at small viewing angles are more sensitive to the model applied for
describing Cherenkov light production.

\end{document}